\documentclass[preprint,preprintnumbers,amsmath,amssymb,superscriptaddress,floatfix]{revtex4}
\usepackage{graphicx}
\usepackage{dcolumn}
\usepackage{bm}
\begin{document}

\title{Spin-orbital liquid in Ba$_3$CuSb$_2$O$_9$ stabilized by oxygen holes}

\author{Kou Takubo}
\email{takubo.k.ab@m.titech.ac.jp}
\affiliation{Institute for Solid State Physics, University of Tokyo, Kashiwa, Chiba 277-8581, Japan}
\affiliation{Department of Chemistry, Tokyo Institute of Technology, Meguro, Tokyo 152-8551, Japan}
\author{Takashi Mizokawa}
\affiliation{Department of Applied Physics, Waseda University, Okubo, Tokyo 277-8581, Japan}
\author{Huiyuan Man}
\affiliation{Institute for Solid State Physics, University of Tokyo, Kashiwa, Chiba 277-8581, Japan}
\affiliation{Institute for Quantum Matter and Department of Physics and Astronomy, Johns Hopkins University, Baltimore, Maryland 21218, USA}
\author{Kohei Yamamoto}
\affiliation{Institute for Solid State Physics, University of Tokyo, Kashiwa, Chiba 277-8581, Japan}
\author{Yujun Zhang}
\affiliation{Institute for Solid State Physics, University of Tokyo, Kashiwa, Chiba 277-8581, Japan}
\author{Yasuyuki Hirata}
\affiliation{Institute for Solid State Physics, University of Tokyo, Kashiwa, Chiba 277-8581, Japan}
\author{Hiroki Wadati}
\affiliation{Institute for Solid State Physics, University of Tokyo, Kashiwa, Chiba 277-8581, Japan}
\affiliation{Graduate School of Material Science, University of Hyogo, Koto, Hyogo 678-1297, Japan}
\author{Akira Yasui}
\affiliation{Japan Synchrotron Radiation Research Institute(JASRI/SPring-8), Sayo, Hyogo 678-1297, Japan}
\author{Daniel I. Khomskii}
\affiliation{II Physikalisches Institut, Universit\"{a}t zu K\"{o}ln, Z\"{u}lpicher Strasse, 50937 K\"{o}ln, Germany}
\author{Satoru Nakatsuji}
\affiliation{Institute for Solid State Physics, University of Tokyo, Kashiwa, Chiba 277-8581, Japan}
\affiliation{Institute for Quantum Matter and Department of Physics and Astronomy, Johns Hopkins University, Baltimore, Maryland 21218, USA}
\affiliation{Department of Physics, University of Tokyo, Hongo, Tokyo 113-0033, Japan}

\date{\today}

\begin{abstract}
Both the Jahn-Teller distortion of Cu$^{2+}$O$_6$ octahedra and magnetic ordering are absent 
in hexagonal Ba$_3$CuSb$_2$O$_9$ suggesting a Cu 3$d$ spin-orbital liquid state.
Here, by means of resonant x-ray scattering and absorption experiment, we show that oxygen 2$p$ holes play
crucial role in stabilizing this spin-orbital liquid state.
These oxygen holes appear due to the ``reaction'' Sb$^{5+}$$\rightarrow$Sb$^{3+}$ $+$ two oxygen holes,
with these holes being able to attach to Cu ions.
The hexagonal phase with oxygen 2$p$ holes exhibits also 
a novel charge-orbital dynamics which is absent in the orthorhombic phase of Ba$_3$CuSb$_2$O$_9$ with Jahn-Teller distortion and Cu 3$d$ orbital order. 
The present work opens up a new avenue towards spin-charge-orbital entangled liquid state in transition-metal oxides with small or negative charge transfer energy. 
\end{abstract}

\maketitle

Mott insulators with $S$=1/2 or $J$=1/2 are expected to show a variety of quantum spin liquid states when magnetic orders are suppressed due to geometrical frustration and/or quantum fluctuation \cite{Kitaev2006,Lee2008,Balents2010}. Among them, RuCl$_3$ with Ru$^{3+}$($J$=1/2) hexagonal lattice exhibits a Kitaev spin liquid state under magnetic field due to Majorana quantization \cite{Do2017,Kasahara2018,Baskaran2007,Jackeli2009,Knolle2014,Nasu2015}. The Kitaev state can be realized by the strong spin-orbit interaction of 4$d$ or 5$d$ transition-metal ions on the hexagonal lattice \cite{Jackeli2009}. Another quantum spin liquid state on a hexagonal lattice is realized in Ba$_3$CuSb$_2$O$_9$ with Cu$^{2+}$($S$=1/2) \cite{Zhou2011,Nakatsuji2012}. The spin liquid phase in Ba$_3$CuSb$_2$O$_9$ can be associated with fluctuation in the orbital sector \cite{Feiner1997,Li1998,Vernay2004,Nasu2008} in contrast to the Kitaev state 
where the orbital degeneracy is lifted due to the strong spin-orbit interaction. 

Ligand holes, e.g. oxygen holes often present in oxides with, typically, high valence (or high oxidation state) of a metal (such as  nominally Fe$^{4+}$ in CaFeO$_3$ \cite{CaFeO3},  Cu$^{3+}$ in NaCuO$_2$ \cite{Mizokawa1991} or Bi$^{4+}$ in BaBiO$_3$ \cite{BaBiO3_1,BaBiO3_2}) lead to a lot of nontrivial effect, which largely determine the properties of corresponding solids \cite{Khomskii,Green}. Among them, novel types of magnetic ordering, like up-up-down-down structure of $R$NiO$_3$ ($R$ - rare earths ions \cite{Mizokawa2000}; spontaneous charge, or rather valence bond disproportionation in CaFeO$_3$ \cite{CaFeO3} and BaBiO$_3$ \cite{BaBiO3_1,BaBiO3_2}, and, last but not least, their apparently fundamental role in High-T$_{\mathrm c}$ superconductivity in cuprates (presumably connected with the formation of Zhang-Rice singlets \cite{ZhangRice}).
Here we uncover yet another nontrivial effect due to the presence or oxygen holes: the suppression of {{{cooperative}}} Jahn-Teller (JT) distortion and magnetic ordering, typical for the strong JT ion Cu$^{2+}$, with the resulting formation of spin-orbital liquid state in a very interesting material Ba$_3$CuSb$_2$O$_9$ \cite{Nakatsuji2012}.
We also show that, besides the suppression of the JT and magnetic ordering, oxygen holes lead to a very specific dynamic effects.

The octahedrally coordinated Cu$^{2+}$ with $3d^9$ electronic configuration 
is known as one of the classical JT active ions.
Usually, divalent CuO$_6$ octahedron in concentrated systems always leads to a cooperative orbital ordering with the concomitant lattice distortion (cooperative JT effect, see e.g. Ref. \cite{KugelKhomskii}).
When the CuO$_6$ octahedron is elongated along the $z$-axis, the Cu 3$d$ orbital with $x^2-y^2$ symmetry is destabilized and accommodates the Cu 3$d$ hole. Ba$_3$CuSb$_2$O$_9$ harbors orthorhombic phase \cite{Zhou2011} and hexagonal phase \cite{Nakatsuji2012,Katayama2015}. 
A very unusual and unexpected effect -- the absence of JT distortion in a classical strong JT ion Cu$^{2+}$ in octahedral coordination was discovered in the hexagonal Ba$_3$CuSb$_2$O$_9$ by Nakatsuji \textit{et al} \cite{Nakatsuji2012}. There are very convincing experimental indications of a dynamic character of the behaviour of Cu in this system \cite{Ishiguro2013, Han2015}. However, the real microscopic origin of this behaviour was not elucidated in these detailed studies.  Just the usual dynamic JT effect can hardly explain this behaviour: it is usually realized in well-isolated JT centers, but never in concentrated systems.  An extra puzzle is that the same material seems to exist in  two different modifications. The orthorhombic phase behaves quite normally: it shows a cooperative JT distortion, no special dynamic effects, etc. On the other hand, the hexagonal phase has all these strange features discussed above. What is the microscopic reason for this very different behaviour, remains a puzzle.
Here, we present new experimental data which shed light on these questions and which give us the possibility to solve all these puzzles. Notably, our spectroscopic investigations show that, whereas the orthorhombic phase contains the usual Cu$^{2+}$ with all the conventional features thereof, in the hexagonal phase we see definite signatures of the presence of substantial oxygen hole character. Fluctuations, inevitably appearing due to motion of these oxygen holes hopping from site to site, suppress conventional long-range JT 
ordering and magnetic ordering and cause the dynamics seen in ESR and NMR \cite{Ishiguro2013, Han2015}.


\begin{figure}[t!]
\includegraphics[width=1\linewidth]{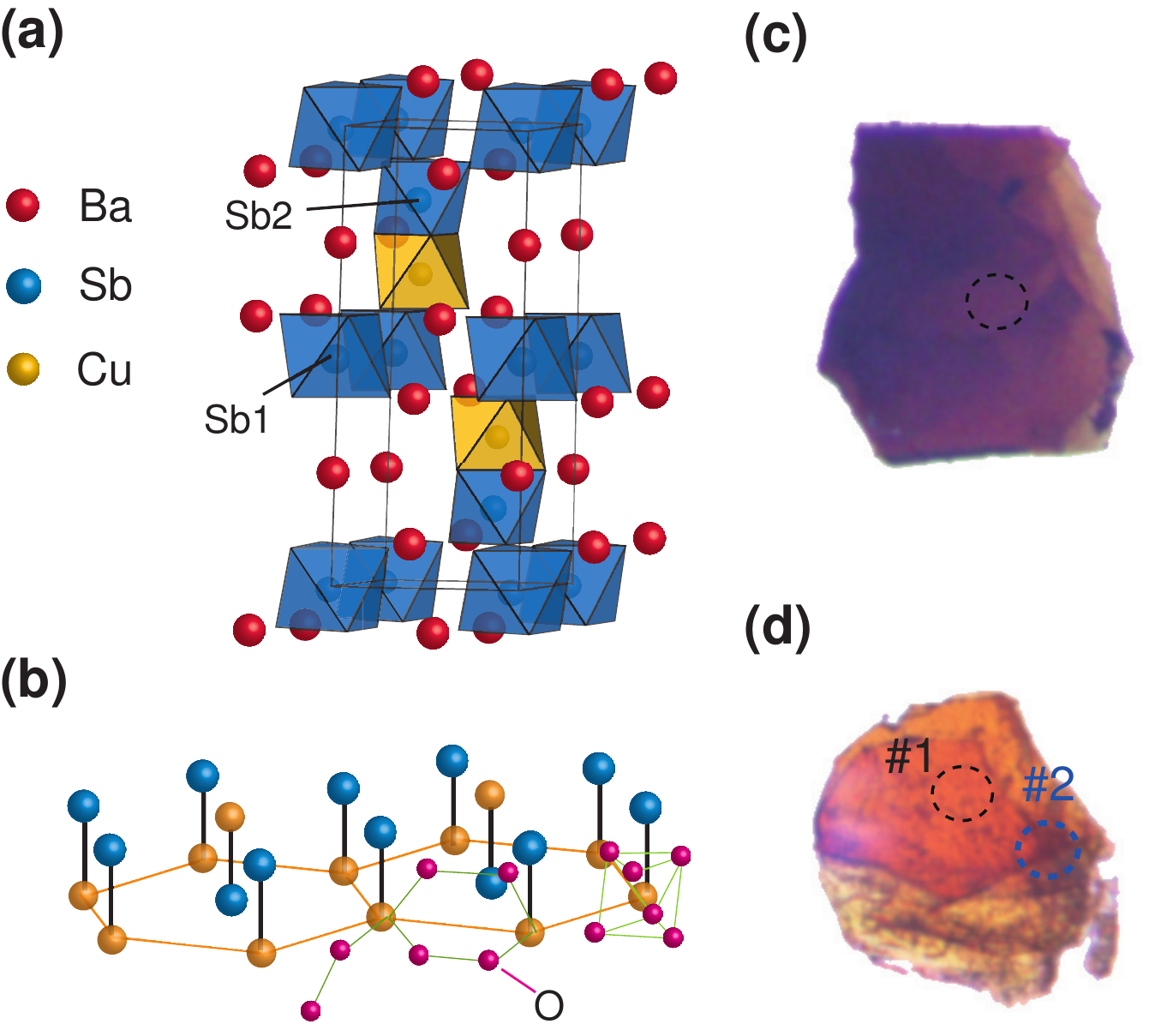}%
\caption{
		Schematic structure and photographs of Ba$_3$CuSb$_2$O$_9$.
		(a,b) Crystal structure. Sb ions and octahedra are shown by blue, and Cu - by yellow.
		The space groups of hexagonal and orthorhombic phases at low temperature are $P6_3/mmc$ and $Cmcm$, respectively \cite{Nakatsuji2012,Katayama2015}.
		The occupancy of Cu and Sb in the two metal sites of the Cu-Sb face-sharing dumbbells is 50 \%.
		The Cu octahedra in the dumbbells form honeycomb lattices in the $ab$-plane as shown in (b), which are isolated and connected by O-O bonds.  
		(c,d) Sample photographs of hexagonal and orthorhombic samples.
		The thicknesses of both samples are below 5 $\mu$m.
		The dashed circles ($\phi\sim$ 50 $\mu$m) correspond to the selected regions for the optical absorption microscopy of Fig. 2(a).
	}
\end{figure}


The basic crystal structure of Ba$_3$CuSb$_2$O$_9$ is shown in Figs. 1(a) and 1(b) (hexagonal phase has symmetry $P6_3/mmc$). Sb ions (blue balls/octahedra) occupy isolated octahedra and (in an ordered way) half of octahedra forming face-sharing dimers \cite{Nakatsuji2012,Katayama2015}. Cu ions (yellow) occupy other half of dimer octahedra, so that they form in effect the honeycomb lattice in the $ab$-plane as shown in Fig. 1(b). The structure of orthorhombic modification (symmetry $Cmcm$) is basically similar, but CuO$_6$ octahedra are strongly distorted due to Jahn-Teller effect typical for Cu$^{2+}$.  

Single crystals of Ba$_3$CuSb$_2$O$_9$ were grown under oxygen atmosphere from the BaCl$_2$-based flux \cite{Katayama2015}.
Two types of single crystals were obtained, depending on the growth condition.
A small addition (9 mol \%) of Ba(OH)$_2$ to the BaCl$_2$ flux was found to stabilize single-phase
crystals of the hexagonal samples, whereas the pure BaCl$_2$ flux leads to the growth of orthorhombic samples.
The composition analysis by ICP-AES indicates that hexagonal samples are a single phase, and stoichiometric in terms of the Cu to Sb elemental ratio \cite{Katayama2015}.
On the other hand, the orthorhombic crystals are rather offstoichiometric and seemed to contain a part of the inhomogeneous domain of the hexagonal phase, as discussed in the optical microscopy measurement.
XAS and time-resolved soft x-ray scattering at the Cu $L$ (2$p\rightarrow$ 3$d$) and O $K$ (1$s\rightarrow$ 2$p$) absorption edges were conducted at BL07LSU in SPring-8 \cite{Takubo2017}.
Further technical details about the optical and x-ray spectroscopies are described in the Supplementary Material \cite{supp}.

An obvious difference between the hexagonal and orthorhombic phases is the 'color' of the crystals. The crystal with dark color [Fig. 1(c)] is dominated by the hexagonal phase while the yellowish color crystal [Fig. 1(d)] mainly contains the orthorhombic phase.
Figure 2(a) shows the optical absorption spectra for the hexagonal and orthorhombic crystals.
The orthorhombic crystals exhibit inhomogeneity. The spectrum taken at position \#1[see Fig. 1(d)] represents the orthorhombic phase with optical gap $h\nu$ $\sim$ 2.0 eV which corresponds to the oxygen 2$p$ to Cu 3$d$ charge transfer excitation as expected in Cu$^{2+}$ oxides. On the other hand, the spectrum at position \#2 exhibits an intriguing absorption centered around 1.3 eV ($\sim$ 950 nm) below the charge transfer excitation $\sim$ 2.0-3.0 eV. The appearance of the absorption peak in the charge transfer gap suggests that some holes are introduced in the Cu$^{2+}$ Mott insulating state, similar to the high-$T_c$ cuprates. Surprisingly, in the absorption spectrum for the hexagonal phase, the in-gap absorption peak around 1.3 eV gains substantial spectral weight which causes the dark color. This indicates that a considerable number of holes are doped in the hexagonal phase although it is highly insulating. 

\begin{figure}[t!]
	\includegraphics[width=1\linewidth]{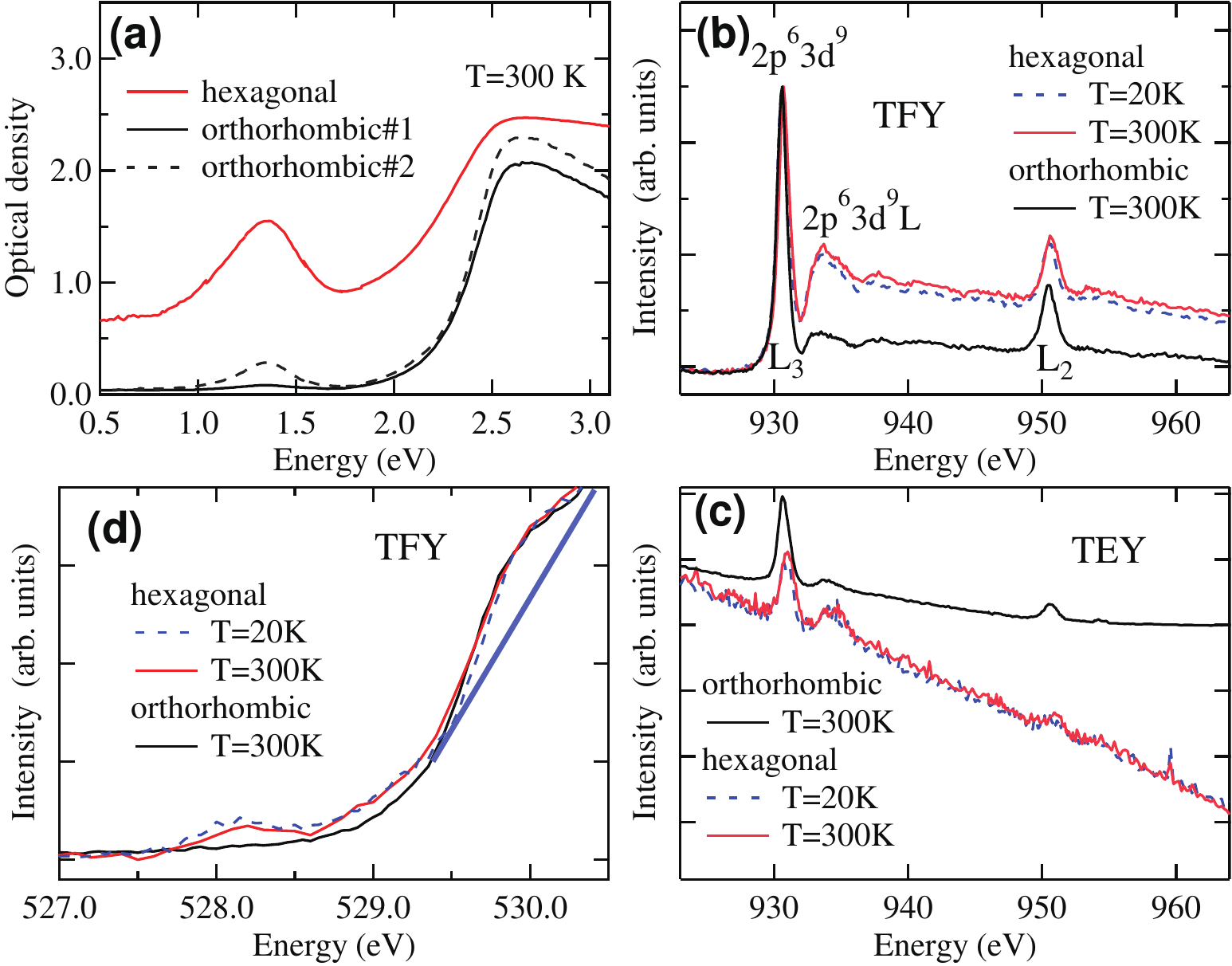}%
	\caption{Differences in the electronic states of the hexagonal and orthorhombic phases of Ba$_3$CuSb$_2$O$_9$.
	(a) Typical optical absorption of the hexagonal and orthorhombic phases estimated from their transmission,
	which are taken in the transmission microscopy mode with space resolution of 30 $\mu m$.  
	(b) Typical Cu 2$p$ XAS spectra of the hexagonal and orthorhombic phases taken in the total fluorescence yield mode.
	(c) Typical Cu 2$p$ XAS spectra of the hexagonal and orthorhombic phases taken in the total electron yield mode.
	(d) Typical O $1s$ XAS spectra of the hexagonal and orthorhombic phases taken in the total fluorescence yield mode.}
\end{figure}

In order to further clarify this electronic structure difference between the hexagonal and orthorhombic phases, x-ray absorption spectroscopy have been investigated for both phases.
Figure 2(b) shows the Cu 2$p$ x-ray absorption spectra (XAS) in the total fluorescence yield mode.
The Cu 2$p$ main peak at 930.2 eV is accompanied by the charge transfer satellite at about 933 eV. The intensity of the satellite is enhanced in the hexagonal phase compared to the orthorhombic phase. The main and satellite peaks are assigned to the transitions of $2p^63d^9$ $\rightarrow$ $2p^53d^{10}$ and of $2p^63d^{9}L$ $\rightarrow$ $2p^53d^{10}L$ \cite{Mizokawa1991}. Here $L$ represents a hole on the oxygen 2$p$ orbitals. The Cu 2$p$ XAS indicates that the ground state of the hexagonal phase include more $3d^{9}L$ than the orthorhombic phase. Figure 2{\it C} shows the Cu 2$p$ XAS 
taken in the total electron yield mode. The XAS spectra in the total electron yield mode are more surface sensitive than those in the total fluorescence yield mode. The XAS results obtained in the two modes are very consistent indicating that the contribution of $3d^9L$ is not due to the surface effect. 
In the hexagonal phase, the intensity of the satellite is comparable to that of the main peak indicting that the oxygen hole concentration is about 1/6 [see Fig. S3(a) in Supplementary Material]\cite{supp}.

Figure 2(d) shows the oxygen 1$s$ XAS spectra of the hexagonal and orthorhombic regions. The structure at 530 eV can be assigned to the transition from 1$s$ to 2$p$
of the oxygen
which is hybridized into the upper Hubbard band (corresponding to $3d^{9}$ $\rightarrow$ $3d^{10}$). The peak at 528.2 eV is observed only in the hexagonal region and can be assigned to the Zhang-Rice singlet band seen in the hole-doped high-$T_c$ cuprates: the transition from oxygen 1$s$ to unoccupied oxygen 2$p$ (corresponding to $3d^{9}L$ $\rightarrow$ $3d^{9}$) \cite{Chen1990}.
The band around 1.3 eV seen in the optical absorption of Fig. 2(a)  corresponds to the excitation from these $3d^{9}L$ state.
The intensity of oxygen hole band (Zhang-Rice singlet band) in the O 1$s$ XAS spectra is comparable to that of the upper Hubbard band after removing the Sb-O contribution which is roughly indicated by the blue line in Fig. 2(d). The comparison with the calculation suggests that the hole concentration per Cu site (or Sb2) site is around 1/6.

To get an extra information on the behaviour of our system, we carried out the time-resolved resonant x-ray scattering at 930.2 eV (on the Cu 2$p$ to 3$d$ resonance) after the pump pulse at 3.1 eV (Fig. 3). 
A coherent oscillation with period of $\sim$ 165 ps is clearly observed in the hexagonal phase, while no oscillation is seen in the orthorhombic phase.
In addition, the x-ray scattering signals probed at 923.0 eV below the absorption edge do not show appreciable change after the pump pulse. 
The Fourier transform of the time-resolved data is plotted in Fig. 3(b) indicating that the coherent oscillation corresponds to $\sim$ 6 GHz which is rather slow compared to various optical phonons in the system which are usually coupled to charge or orbital orderings. Such a slow dynamics in the hexagonal phase is consistent with the previous reports \cite{Ishiguro2013, Han2015}.

\begin{figure}[t!]
	\includegraphics[width=1\linewidth]{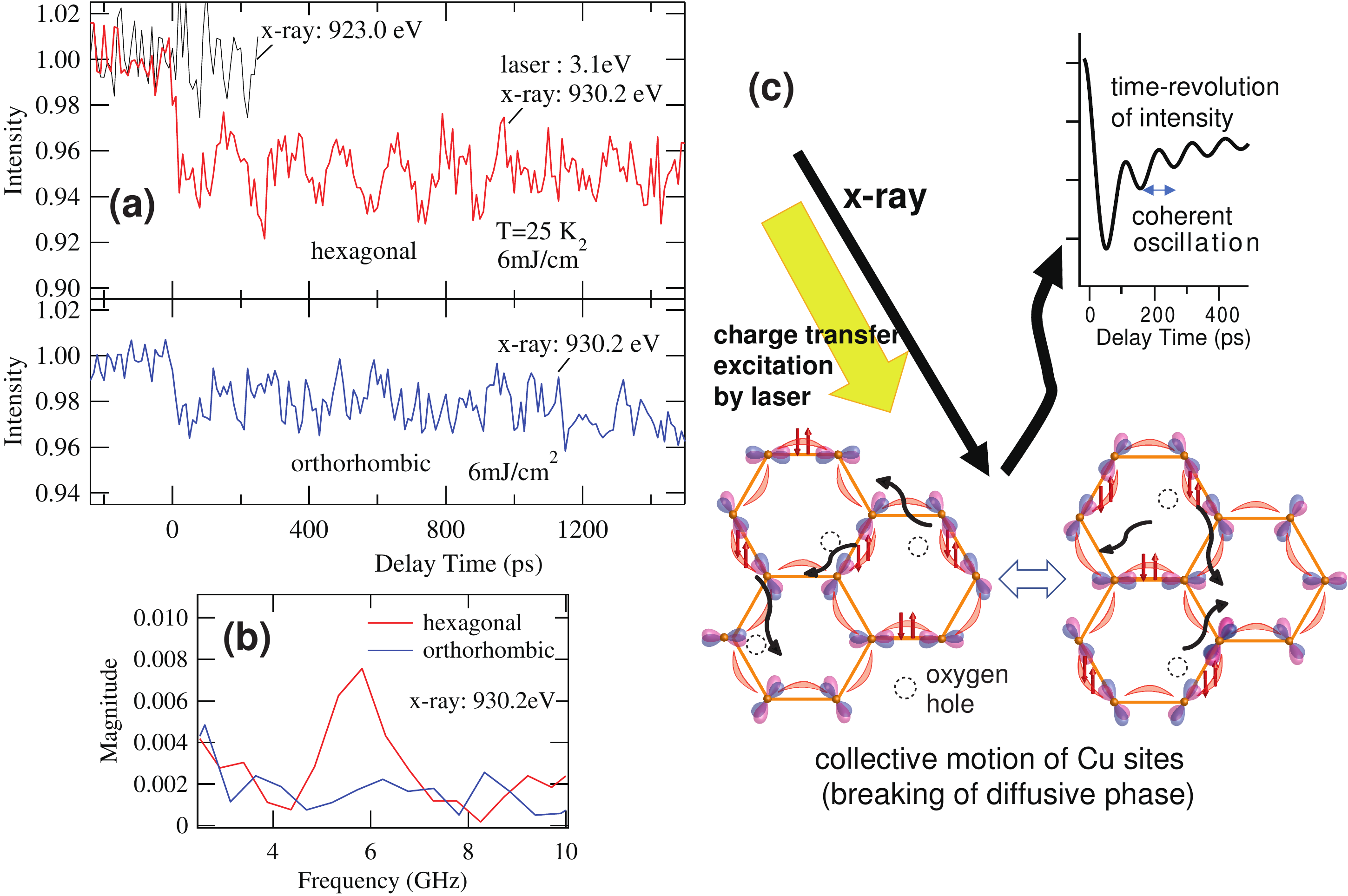}%
	\caption{Time-resolved resonant x-ray scattering of Ba$_3$CuSb$_2$O$_9$.
	(a) Photo-induced dynamics at $\rm Q$=(002) with x-ray of 930.2 eV (and 923.0 eV) for the hexagonal and orthorhombic phases after the pump pulse at 3.1 eV.
	(b) Fourier transform of the time-resolved resonant x-ray scattering signals.
	(c) Schematic picture of photo-induced charge-orbital dynamics in the hexagonal phase.
}
\end{figure}

The presence of the oxygen 2$p$ hole in the hexagonal phase indicates that the Cu 3$d$ spin and orbital are disturbed by hopping of the oxygen 2$p$ holes. 
Since the hexagonal system with oxygen holes remains highly insulating, the oxygen hole should be confined within several Cu sites and disturbs the Cu 3$d$ spin and orbital of those Cu sites. One of the possible units of confinement is the hexagonal cluster shown in Fig. 3(c) where the six CuO$_6$ octahedra are connected through the Cu-O-O-Cu pathways. In this cluster, the oxygen hole and Cu 3$d$ spins/orbitals form a quantum object keeping the hexagonal symmetry. As for the coherent oscillation seen in the time-resolved x-ray scattering in hexagonal samples, their source could be motion of oxygen holes in such clusters. 
The coherent oscillation with period of $\sim$165 ps is absent in the orthorhombic sample as shown in Fig. 3(b). Since the pump pulse with 3.1 eV corresponds to the charge transfer excitation from the O 2$p$ to Cu 3$d$ orbitals, the coherent oscillation is related to the charge dynamics in the Cu-O-O-Cu network. The most significant difference between the hexagonal and orthorhombic samples is the presence of oxygen hole in the hexagonal sample. Therefore, it is natural to assign the slow dynamics of $\sim$165 ps is derived from the oxygen hole in the Cu-O-O-Cu network.

There are several possible origins of the oxygen holes in Ba$_3$CuSb$_2$O$_9$.
The most plausible possibility is the high valence Sb in the Cu-Sb dimers. %
The formal valence is Cu$^{2+}$ and Sb$^{5+}$. 
However, the valence state 5+ is rather high for Sb. For such high oxidation states the real electronic configuration tends to contain ligand holes. For example, in BaBiO$_3$ with Bi$^{5+}$/Bi$^{3+}$ mixed valence, it is established that the formally Bi$^{5+}$ site has the actual electronic configuration close to Bi$^{3+}L^2$ \cite{Foyevtsova2015}. The same trend is expected for Sb which is located above Bi in the periodic table. Indeed, the actual valence of Sb is rather close to 3+ as indicated by the HAXPES result presented in Supplementary Material \cite{supp}. Therefore, the Sb 5$s$ orbitals 'grab' some electrons from the oxygen 2$p$ orbitals and create oxygen 2$p$ holes. If the oxygen 2$p$ holes are bounded to the Sb site, the electronic configuration in the Cu-Sb dimer is essentially the same as Cu$^{2+}$-Sb$^{5+}$, i.e. we would have cluster Cu$^{2+}$[Sb$^{3+}L^2$] with ligand holes forming something like Zhang-Rice state with Sb. However, this ligand hole can be also attached to Cu, making the state [Cu$^{2+}L$][Sb$^{3+}L$] or even [Cu$^{2+}L$]$_2$Sb$^{3+}$. 
In the simple schematic picture oxygen holes may be thus treated as  moving in a double-well potential, one well centered on Cu and another, deeper well - on Sb, with the barrier in between.
The average Sb-O bond length is about 2.004 {\AA} for the Sb2 site while it is about 1.99 {\AA}  for the Sb1 site \cite{Katayama2015}. Consequently, the bond valence sum \cite{Palenik} for the Sb1 and Sb2 sites are estimated to be 4.81 and  4.35, respectively. Therefore, the Sb2 site accommodates the extra electron to create the oxygen hole. 


Fluctuation of ligand holes attached either to Sb or to Cu could be the reason of suppression of the long-range Jahn-Teller ordering of Cu in hexagonal phase. This is the main physics we deduce from our new experimental data, which, in our opinion, can explain the main features of the unexpected behavior of Ba$_3$CuSb$_2$O$_9$. But even regardless of the origin of the oxygen hole, the present results rigorously prove that the oxygen hole and the Cu spins form the unique quantum object with spin-charge-orbital fluctuations of the specific time scale in the hexagonal phase. The Cu deficiency or the partial substitution of Cu by Sb was reported for the orthorhombic phase \cite{Katayama2015}. If one Cu$^{2+}$ is replaced by Sb$^{5+}$, three ligand holes can be eliminated. Therefore, the ligand holes are removed from the system due to the off-stoichiometry, and the Jahn-Teller distortion is recovered in the orthorhombic phase. This picture is indeed consistent with the O 1$s$ XAS spectra in Fig. 2.

It is known that oxygen holes play significant roles in high valence transition-metal oxides such as Cu$^{3+}$, Ni$^{3+}$, and Fe$^{4+}$ oxides. High valence transition-metal oxides with 90 degrees $M$-O-$M$ bonds
are insulating since the oxygen 2$p$ hole mixed with the $M$ 3$d$ $e_g$ hole in a $M$O$_4$ or $M$O$_6$ cluster cannot hop to neighboring clusters and is confined in the single-site cluster \cite{Mizokawa1991}.
On the other hand, high valence transition-metal oxides with almost 180 degrees $M$-O-$M$ bonds
are metallic and their oxygen 2$p$ holes are highly itinerant. With decreasing the $M$-O-$M$ bond angle in $R$NiO$_3$ ($R$ = rare earth metal) or $A$FeO$_3$ ($A$ = alkaline earth metal) by reducing  radius of the $R$ or $A$ ions, oxygen holes and transition-metal spins are ordered and form spin-charge ordered insulating states \cite{Mizokawa2000,Bisogni2016}.
Compared to these other oxygen hole states, 
the oxygen hole state in Ba$_3$CuSb$_2$O$_9$ is very unique in that 
due to a specific crystal structure with rather isolated CuO$_6$ 
octahedra, the ligand holes have weaker tendency to become 
itinerant (even with large concentration of ligand holes the 
hexagonal samples of  Ba$_3$CuSb$_2$O$_9$ remain highly insulating).
This can also lead to a significant difference of XAS spectra in 
this system as compared e.g. with the high-$T_c$ cuprates, see
the discussion in the Supplementary Material \cite{supp}.


In conclusion, x-ray and optical absorption experiments on Ba$_3$CuSb$_2$O$_9$ reveal that the hexagonal phase with spin and orbital liquid is characterized by emergence of oxygen 2$p$ holes in the highly insulating state. Their appearance  originates due to tendency of Sb$^{5+}$ to change electronic state by creating oxygen holes. Oxygen holes may fluctuate between Sb and Cu ions, or may be confined in the relatively large hexagonal cluster with six Cu sites, and the CuO$_6$ JT distortion and long-range magnetic ordering are suppressed due to hopping of oxygen holes between Cu and Sb and between CuO$_6$ octahedra. This can be a rather general situation in many materials containing different metal ions with relatively high valence, in which ligand (e.g. oxygen) holes can play crucial role in determining their properties. And the method we used  - x-ray spectroscopy - can be one of the best methods to unravel interesting physics connected with these phenomena. Our results demonstrate strong interplay between electronic structure with the presence of oxygen holes in systems with unusual valencies, and the Jahn-Teller effect - the effect which often plays crucial role in concentrated solids including high-$T_\mathrm{c}$ superconducting cuprates, but also in molecular systems and in inorganic chemistry.
The coherent oscillations of 6 GHz, clearly observed in pump-probe resonant x-ray scattering, indicate that the spin and orbital fluctuations in the hexagonal phase have relatively slow dynamics, and suggests that the spin-charge-orbital fluctuation can be controlled by optical excitation. The present results also pave a new avenue towards optical control of the spin-charge-orbital states in transition-metal compounds with rich physical properties.

\section*{Acknowledgements}

We are grateful for fruitful discussions with and support from S. Koshihara and T. Ishikawa. XAS and Tr-RSXS measurements were performed under the approval of Synchrotron Radiation Research Organization, the University of Tokyo (No. 2015A7401, 2018B7577). HAXPES measurements were performed under the approval of SPring-8 (Proposals No. Proposal No. 2018B1449). The work of D.Kh. is funded by the Deutsche Forschungsgemeinschaft (DFG, 
German Research Foundation) - Project number 277146847 - CRC 1238.




\begin{references}
		
\bibitem{Kitaev2006}
A. Kitaev, {\it Anyons in an exactly solved model and beyond,} Ann. Phys. {\bf 321}, 2-111 (2006).

\bibitem{Lee2008}
P. A. Lee, Physics. {\it An end to the drought of quantum spin liquids,} Science {\bf 321}, 1306-1307 (2008).

\bibitem{Balents2010}
L. Balents, {\it Spin liquids in frustrated magnets,} Nature {\bf 464}, 199-208 (2010).

\bibitem{Do2017}
S.-H. Do, S.-Y. Park, J. Yoshitake, J. Nasu, Y. Motome, Y. S. Kwon, D. T. Adroja, D. J. Voneshen, K. Kim, T.-H. Jang, J.-H. Park, K.-Y. Choi, and S. Ji, {\it Majorana fermions in the Kitaev quantum spin system $\alpha$-RuCl$_3$,} Nature Physics {\bf 13}, 1079-1084 (2017).

\bibitem{Kasahara2018}
Y. Kasahara, T. Ohnishi, Y. Mizukami, O. Tanaka, Sixiao Ma, K. Sugii, N. Kurita, H. Tanaka, J. Nasu, Y. Motome, T. Shibauchi, and Y. Matsuda, {\it Majorana quantization and half-integer thermal quantum Hall effect in a Kitaev spin liquid,} Nature {\bf 559}, 227-231 (2018).

\bibitem{Baskaran2007}
G. Baskaran, S. Mandal, and R. Shankar, {\it Exact results for spin dynamics and fractionalization in the Kitaev model,} Phys. Rev. Lett. {\bf 98}, 247201 (2007).

\bibitem{Jackeli2009}
G. Jackeli, and G. Khaliullin, {\it Mott Insulators in the Strong Spin-Orbit Coupling Limit: From Heisenberg to a Quantum Compass and Kitaev Models,} Phys. Rev. Lett. {\bf 102}, 017205 (2009).

\bibitem{Knolle2014}
J. Knolle, D. L. Kovrizhin, J. T. Chalker, and R. Moessner, {\it Dynamics of a two-dimensional quantum spin liquid: signatures of emergent Majorana fermions and fluxes,} Phys. Rev. Lett. {\bf 112}, 207203 (2014).

\bibitem{Nasu2015}
J. Nasu, M. Udagawa, and Y. Motome, {\it Thermal fractionalization of quantum spins in a Kitaev model: temperature-linear specific heat and coherent transport of Majorana fermions,} Phys. Rev. B {\bf 92}, 115122 (2015).


\bibitem{Zhou2011}
H. D. Zhou, E. S. Choi, G. Li, L. Balicas, C. R. Wiebe, Y. Qiu, J. R. D. Copley, and J. S. Gardner, {\it Spin liquid state in the S = 1/2 triangular lattice Ba$_3$CuSb$_2$O$_9$,}
Phys. Rev. Lett. {\bf 106}, 147204 (2011).

\bibitem{Nakatsuji2012}
S. Nakatsuji, K. Kuga, K. Kimura, R. Satake, N. Katayama, E. Nishibori, H. Sawa, R. Ishii, M. Hagiwara, F. Bridges, T. U. Ito, W. Higemoto, Y. Karaki, M. Halim, A. A. Nugroho, J. A. Rodriguez-Rivera, M. A. Green, and C. Broholm, {\it Spin-Orbital Short-Range Order on a Honeycomb-Based Lattice,} Science {\bf 336}, 559-563 (2012).


\bibitem{Feiner1997}
L. F. Feiner, A. M. Oles, and J. Zaanen, {\it Quantum melting of magnetic order due to orbital fluctuations,} Phys. Rev. Lett. {\bf 78}, 2799 (1997).

\bibitem{Li1998}
Y. Li, Q. Ma, M. D. N. Shi, and F. C. Zhang, {\it SU(4) theory for spin systems with orbital degeneracy,} Phys. Rev. Lett. {\bf 81}, 3527 (1998).

\bibitem{Vernay2004}
F. Vernay, K. Penc, P. Fazekas, and F. Mila, {\it Orbital degeneracy as a source of frustration in LiNiO$_2$,} Phys. Rev. B {\bf 70}, 014428 (2004).

\bibitem{Nasu2008}
J. Nasu, A. Nagano, M. Naka, and S. Ishihara, {\it Doubly degenerate orbital system in honeycomb lattice: Implication of orbital state in layered iron oxide,} Phys. Rev. B {\bf 78}, 024416 (2008).

\bibitem{CaFeO3}
M. Takano, S. Nasu, T. Abe, K. Yamamoto, S. Endo, Y. Takeda, and J. B. Goodenough, {\it Pressure-inuced high-spin to low-spin transition in CaFeO$_3$,} Phys. Rev. Lett. {\bf 67}, 3267 (1991).

\bibitem{Mizokawa1991}
T. Mizokawa, H. Namatame, A. Fujimori, K. Akeyama, H. Kondoh, H. Kuroda, and N. Kosugi, {\it Origin of the band gap in the negative charge-transfer-energy compound NaCuO$_2$,} Phys. Rev. Lett. {\bf 67}, 1638 (1991).

\bibitem{BaBiO3_1} T. M. Rice, and L. Sneddon, {\it Real-Space and $\vec{k}$-Space Electron Pairing in BaPb$_{1-x}$Bi$_x$O$_3$,} Phys. Rev. Lett. {\bf47}, 689 (1981).

\bibitem{BaBiO3_2} C. M. Varma, {\it Missing valence states, diamagnetic insulators, and superconductors,} Phys. Rev. Lett. {\bf61}, 2713 (1988).

\bibitem{Khomskii}
D. I. Khomskii, {\it Transition metal compounds.} (Cambridge Univ.Press, 2014).

\bibitem{Green}
G. A. Sawatzky, and R. Green, "The Explicit Role of Anion States in High-Valence Metal Oxides." in
{\it Quantum Materials: Experiments and Theory Modeling and Simulation Vol. 6}, E. Pavarini, E. Koch, J. van den Brink, and G. A. Sawatzky, Eds., (Verlag des Forschungszentrum, Julich, 2016) pp. 1–36.

\bibitem{Mizokawa2000}
T. Mizokawa, D. I. Khomskii, and G. A. Sawatzky, {\it Spin and charge ordering in self-doped Mott insulators,} Phys. Rev. B {\bf 61}, 11263 (2000).

\bibitem{ZhangRice} F. C. Zhang and T. M. Rice, {\it Effective Hamiltonian for the superconducting Cu oxides,} Phys. Rev. B {\bf37}, 3759(R) (1988).

\bibitem{KugelKhomskii}	
K. I. Kugel' and D. I. Khomskii, {\it The Jahn-Teller effect and magnetism: transition metal compounds,} Sov. Phys. Usp. {\bf25}, 231-256 (1982).


\bibitem{Katayama2015}
N. Katayama, K. Kimura, Y. Han, J. Nasu, N. Drichko, Y. Nakanishi, M. Halim, Y. Ishiguro, R. Satake, E. Nishibori, M. Yoshizawa, T. Nakano, Y. Nozue, Y. Wakabayashi, S. Ishihara, M. Hagiwara, H. Sawa, and S. Nakatsuji, {\it Absence of Jahn-Teller transition in the hexagonal Ba$_3$CuSb$_2$O$_9$ single crystal,} Proc. Natl. Acad. Sci. U.S.A. {\bf 112}, 9305-9309 (2015).

\bibitem{Ishiguro2013}
Y. Ishiguro, K. Kimura, S. Nakatsuji, S. Tsutsui, A. Q. R. Baron, T. Kimura, and Y. Wakabayashi, {\it Dynamical spin-orbital correlation in the frustrated magnet Ba$_3$CuSb$_2$O$_9$,} Nature Comm. {\bf 4}, 2022 (2013).

\bibitem{Han2015}
Y. Han, M. Hagiwara, T. Nakano, Y. Nozue, K. Kimura, M. Halim, and S. Nakatsuji, {\it Observation of the orbital quantum dynamics in the spin-1/2 hexagonal antiferromagnet Ba$_3$CuSb$_2$O$_9$,} Phys. Rev. B {\bf 92}, 180410(R) (2015). 

\bibitem{Takubo2017}
K. Takubo, K. Yamamoto, Y. Hirata, Y. Yokoyama, Y. Kubota, S. Yamamoto, S. Yamamoto, I. Matsuda, S. Shin, T. Seki, K. Takanashi, H. Wadati, {\it Capturing ultrafast magnetic dynamics by time-resolved soft x-ray magnetic circular dichroism,} Appl. Phys. Lett. {\bf 110}, 162401 (2017).

\bibitem{supp}See Supplemental Material at http://link.aps.org/supplemental/ for additional experimental methods, data, and calculations.

\bibitem{Chen1990}
C. T. Chen, F. Sette, Y. Ma, M. S. Hybertsen, E. B. Stechel, W. M. C. Foulkes, M. Schulter, S-W. Cheong, A. S. Cooper, L. W. Rupp, Jr., B. Batlogg, Y. L. Soo, Z. H. Ming, A. Krol, and Y. H. Kao, {\it Electronic states in La$_{2-x}$Sr$_x$CuO$_{4+\delta}$. probed by soft-x-ray absorption,} Phys. Rev. Lett. {\bf 66}, 104 (1991).

\bibitem{Foyevtsova2015}
K. Foyevtsova,  A. Khazraie, I. Elfimov, and G. A. Sawatzky, {\it Hybridization effects and bond disproportionation in the bismuth perovskites,} Phys. Rev. B {\bf 91}, 121114 (2015).






\bibitem{Bisogni2016}
V. Bisogni, S. Catalano, R. J. Green, M. Gibert, R. Scherwitzl, Y. Huang, V. N. Strocov, P. Zubko, S. Balandeh, J.-M. Triscone, G. Sawatzky, T. Schmitt, {\it Ground-state oxygen holes and the metal-insulator transition in the negative charge-transfer rare-earth nickelates,} Nature Commun. {\bf 7}, 13017 (2016).


\bibitem{Palenik}
R.C. Palenik, K. A. Abboud and G. J. Palenik, Inorg. Chim. Acta {\bf 358}, 1034 (2005).

\bibitem{NIST2012}
NIST X-ray Photoelectron Spectroscopy Database, Version 4.1 (National Institute of Standards and Technology, Gaithersburg, 2012); http://srdata.nist.gov/xps/; DOI: http://dx.doi.org/10.18434/T4T88K.

\end{references}
\end{document}


\begin{center}
	\bf	\large
	Supplementary Material for:
	
	\vspace{3mm}
	\bf	\Large	
Spin-orbital liquid in Ba$_3$CuSb$_2$O$_9$ stabilized by oxygen holes

	\vspace{5mm}
	
\normalsize\rm
{K. Takubo,$^{1,2,\ast}$, T. Mizokawa,$^{3}$ H. Man,$^{1,4}$ K. Yamamoto,$^{1}$ Y. Zhang,$^{1,5}$ Y. Hirata,$^{1}$
	\\ H. Wadati,$^{1,5}$ A. Yasui,$^{6}$  D. I. Khomskii,$^{7}$  S. Nakatsuji,$^{1,4,8}$ \\
\vspace{4mm}

{\it
	{\small$^{1}$ Institute for Solid State Physics, University of Tokyo, Kashiwa, Chiba 277-8581, Japan}\\
	{\small$^{2}$ Department of Chemistry, Tokyo Institute of Technology, Meguro, Tokyo 152-8551, Japan}\\
	{\small$^{3}$ Department of Applied Physics, Waseda University, Shinjyuku, Tokyo 169-8555, Japan}\\
	{\small$^{4}$ Institute for Quantum Matter and Department of Physics and Astronomy, Johns Hopkins University, Baltimore, Maryland 21218, USA}\\
	{\small$^{5}$ Graduate School of Material Science, University of Hyogo, Sayo, Hyogo 678-1297, Japan}\\
	{\small{$^{6}$Japan Synchrotron Radiation Research Institute (JASRI/SPring-8), Sayo, Hyogo 679-5198, Japan}\\
	{}\small$^{7}$II Physikalisches Institut, Universit\"{a}t zu K\"{o}ln, Z\"{u}lpicher Strasse, 50937 K\"{o}ln, Germany, Japan}\\
	{\small$^{8}$Department of Physics, University of Tokyo, Hongo, Tokyo 113-0033, Japan}\\
}
	\vspace{2mm}	
	{\small$^\ast$To whom correspondence should be addressed; E-mail: takubo.k.ab@m.titech.ac.jp}
}

\end{center}


\subsection*{Methods}

A Fourier transform infrared spectroscopy microscope (JASCO IRT-30) was used for the optical microscopic measurements in the transmission mode. The space resolution was $\sim$ 30 $\mu$m. The optical absorption was estimated from the transmission.
A typical sample sizes were 150 $\times$ 150 $\times$ 5 $\mu$m$^3$.
XAS and time-resolved soft x-ray scattering at the Cu $L$ (2$p\rightarrow$ 3$d$) and O $K$ (1$s\rightarrow$ 2$p$) absorption edges were conducted at BL07LSU in SPring-8.
The samples were cleaved along the (001) plane \textit{in situ} to avoid the surface contamination.
The XAS spectra were recorded both using the surface-sensitive total electron yield (TEY) and bulk-sensitive total fluorescence yield (TFY) modes.
The time-resolved-RSXS measurements were performed using the pump-probe technique with a time-resolution of $\sim$50 ps \cite{Takubo2017}.
A second-harmonic Ti:sapphire laser pulse ($hv$=3.1 eV, repetition rate = $\sim$1 kHz, width = 50 fs) was adopted
as the pump light.
As a probe light, a synchrotron soft x-ray pulse (width = $\sim$50 ps) with energy near Cu $L_3$ edge was used.
The incident x-ray was sigma-polarized and parallel to the in-plane (100) crystal axis.
The spot size of the x-rays and Ti:sapphire laser is $\sim$100 $\times$ 50 $\mu$m$^2$ and $\sim$600 $\times$ 600 $\mu$m$^2$, respectively. As the area irradiated by synchrotron x-ray is fully exposed by the laser irradiation, the dynamics induced by the laser incidence were probed.
By varying the delay time between the x-ray and the laser pulse, time-resolved information were obtained.
Hard x-ray photoemission spectroscopy (HAXPES) were performed at SPring-8 BL47XU. The photon energy was 7940 eV.
The binding energy was calibrated by the Fermi edge of gold and the energy resolution was approximately 200 meV.

\subsection*{HAXPES measurements}

To evaluate the valence of the Sb sites, we have carried out the hard x-ray photoemission spectroscopy (HAXPES).
Figure S1 shows the HAXPES spectra of the hexagonal Ba$_3$CuSb$_2$O$_9$.
The peak energies of the Sb 3$d_{5/2}$ and 3$d_{3/2}$ are $\sim$529.9 and $\sim$539.3 eV, respectively,
while the Sb 3$d_{5/2}$ peak is just overlapped on the O 1$s$ peak and thus its energy has uncertainty.
These energies are apparently lower than those for the typical Sb${^{5+}}$ (or Sb${^{4+}}$) compounds and similar to Sb${^{3+}}$ as indicated by the dashed lines in the Fig. S1(a).   
The energies of Sb 3$d_{3/2}$, reported on the NIST database including those for Sb$_2$O$_3$, Sb$_2$O$_4$, Sb$_2$O$_5$, and USb$_2$O$_5$\cite{NIST2012}, are ranging from 539.1 to 540.1 eV for 3+, and 540.1 to 540.8 eV for 5+ (or 4+), respectively.
On the other hand, the energies of Sb 3$d_{5/2}$ are ranging from 529.2 to 530.1 eV for 3+, and 530.8 to 532.1 eV for 5+ (or 4+), respectively.
Therefore, these results support the scenario of the Sb$^{3+}$/Sb$^{5+}$ mixed valence in the hexagonal phase of  Ba$_3$CuSb$_2$O$_9$.
The Sb$^{3+}$ sites share the bonding electrons of the oxygens between the Cu-Sb dumbbell like Zhang-Rice singlet state of high-$T_c$ cuprates, and can act as the donor of the holes as discussed in the main text.   

On the other hand, the quantitative estimation of the valence for the Cu sites was hard from the HAXPES spectrum,  since the broad Sb 3$s$ peak was overlapped onto the satellite structure for the Cu 2$p_{3/2}$ observed between 940 and 950 eV [See Fig, S1(b)].
However, the large and sharp satellite structure for Cu 2$p_{1/2}$ observed around 962 eV \cite{Mizokawa1991} is still consistent with the scenario of the oxygen hole in the hexagonal phase.
It should be noted that we had also tried to measure HAXPES for 3 pieces of the orthorhombic crystals,
but all of them were charged up then no meaningful spectrum was obtained.
This fact is also consistent with the highly insulating (or wide-gapped) behaviour of the orthorhombic phase compared to the hexagonal one in the optical spectroscopy. 

\subsection*{Temperature dependence of the resonant x-ray scattering at Cu $L_3$ edge}

The reciprocal Q-vector of (002) was chosen for the time-resolved resonant soft x-ray scattering (RSXS) experiments shown in the main text. Q=(002) is structurally allowed (or a kind of diffraction), and can be accessed in the soft x-ray energy region (2$\theta\sim$ 135$^\circ$ at $hv\sim$ 930 eV) owing to the long $c$-axis $\sim$15.6 {\AA} of Ba$_3$CuSb$_2$O$_9$.
Figure S2 shows the static RSXS spectra around the Cu $L_3$ edge (2$p_{3/2}$ to 3$d$ resonance).
Here, the polarization of the incident x-ray was parallel to (100) of Ba$_3$CuSb$_2$O$_9$.
The RSXS intensity can generally be formulated as $I={|S(\omega)|^2}/\mu(\omega)$, where $S(\omega)$ is the structure factor and given by $S(\omega)=\sum f(\omega)e^{-i{\mathbf Q}\cdot{\mathbf r}}$. Here, $\mu(\omega)$, $f(\omega)$, and $e^{-i{\mathbf Q}{\cdot\mathbf r}}$ are the absorption coefficient, complex dielectric permittivity, and structural component, respectively. 
Because the $c$-axis is unrelated direction for the cooperative distortions of the Cu sites,
the temperature dependence of RSXS at Q=(002) is barely observed below the Cu $L_3$ resonance of $hv$ $<$ 928 eV as shown in Fig. S2(b).
On the other hand, the large spectral change is observable around the Cu $L_3$ resonance of $hv\sim$ 930.2 eV,
reflecting changes of the in-plane dielectric function along (100) of the Cu sites [or $\mu(\omega)$ and $f(\omega)$].
Previous non-resonant in-plane x-ray diffraction (or diffusive scattering) experiments of Ba$_3$CuSb$_2$O$_9$ by Ishiguro \textit{et al}. indicated that ferro- and antiferro-orbital fluctuations develop with cooling \cite{Ishiguro2013}.
The structurally diffusive states are formed at low temperature and its ferro-orbital correlation saturates below the spin-singlet formation temperature of $\sim$ 50 K.
The temperature dependences of the diffusive scattering around Q=(220) [Fig. 3(e) in Ref. \cite{Ishiguro2013}] and RSXS at 930.2 eV given in Fig. S2(c) seem to have a similar trend.
The (time-resolved) RSXS signals probed at 930.2 eV could extract these cooperative changes on the electronic states of the Cu sites, whereas the structural component would less affect the observed changes in RSXS.

\begin{figure}
	\centering
	\includegraphics[width=1\linewidth]{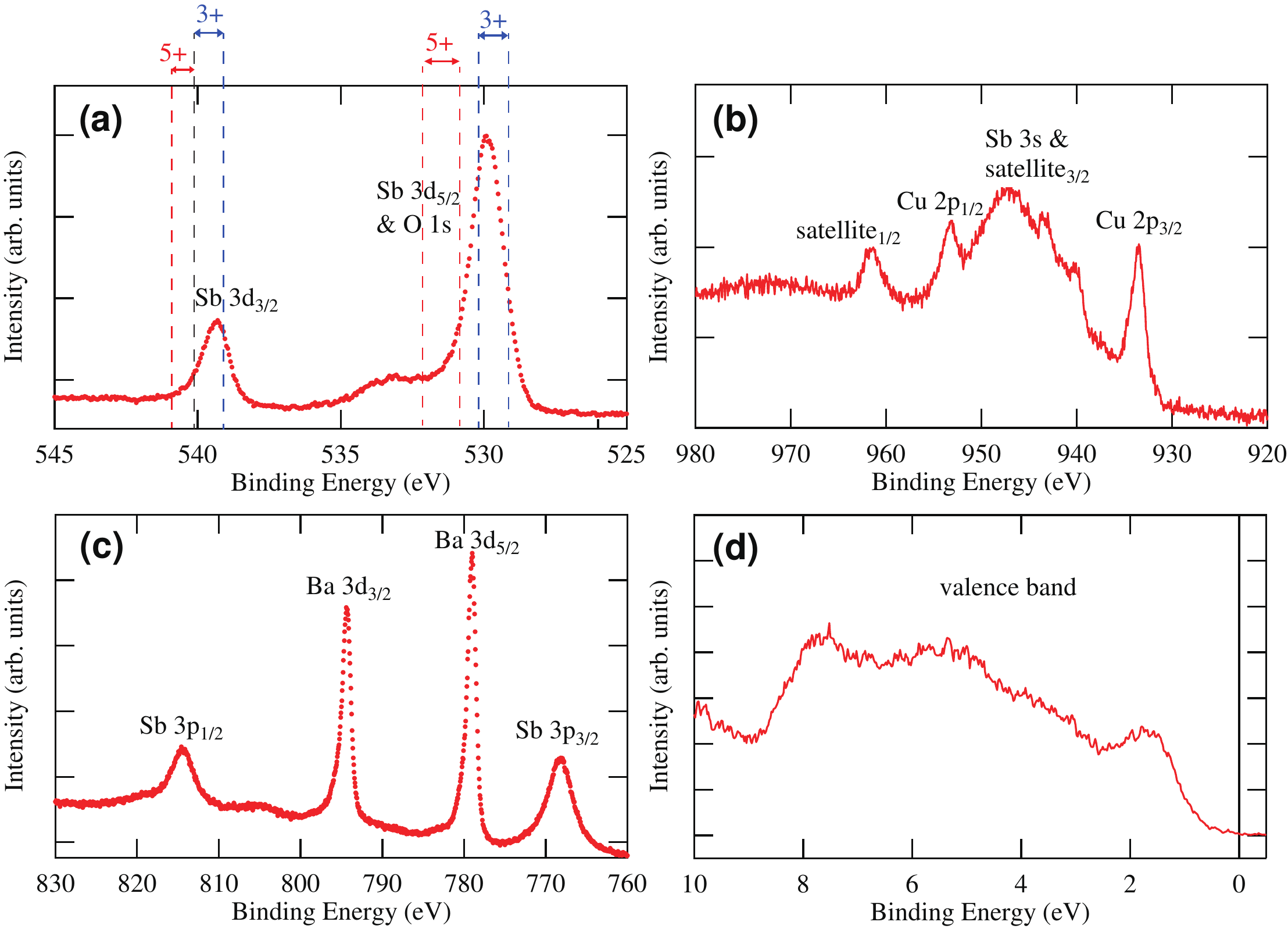}%
	\caption{
		HAXPES spectra of Ba$_3$CuSb$_2$O$_9$ of the hexagonal phase at room temperature.
		(a) Sb 3$d_{5/2,3/2}$ and O 1$s$ spectrum.
		The dashed lines indicate the ranges of the binding energy for typical Sb$^{3+}$ and Sb$^{5+}$ compounds
		reported on the NIST database \cite{NIST2012}.
		(b) Cu 2$p_{3/2,1/2}$ and Sb 3$s$ spectrum.
		(c) Sb 3$p_{3/2,1/2}$ and Ba 3$d_{5/2,3/2}$ spectrum. (d) Valence band spectrum.
	}
\end{figure}

\clearpage

\subsection*{Configuration interaction calculation of XAS spectra}
In Ba$_3$CuSb$_2$O$_9$, the CuO$_6$ octahedra do not share oxygens, while they are directly connected sharing oxygens at their corners in perovskite or high-$T_c$ cuprates. The oxygen holes in the perovskite are shared by two CuO$_6$ octahedra and become itinerant for relatively low hole concentration, providing the canonical behavior of Cu 2$p$ and O 1$s$ x-ray absorption spectra of doped cuprates.
On the other hand, the oxygen hole in Ba$_3$CuSb$_2$O$_9$ can be localized as discussed in the main text, and its Cu 2$p$ and O 1$s$ x-ray absorption spectra exhibit different behaviors.

In order to verify this point, configuration interaction calculation \cite{Mizokawa1991} has been performed for a Cu$_6$O$_{36}$ cluster in which six CuO$_6$ octahedra are connected by O-O bond and have one oxygen hole.
In the cluster, the six CuO$_6$ octahedra with Cu 3$d$ $x^2$-$y^2$ orbital and O 2$p_\sigma$ orbitals are considered.
The calculated results are shown in Fig. S3.
Here, $U_{dd}$ and $U_{cd}$ are Coulomb interaction between 3$d$ states and between the Cu 2$p$ core hole and the Cu 3$d$ hole, respectively.
Delta is the O 2$p$-to-Cu 3$d$ charge transfer energy, and $pd\sigma$ and $pp\sigma$ are the Slater-Koster parameters.
The effect of O 1$s$ core hole potential is neglected.
The transfer integral between the neighboring CuO$_6$ octahedra is given by $(pp\sigma-pp\pi)/\sqrt{2}$ \cite{Nakatsuji2012} which is evaluated assuming $pp\pi  = -1/4 pp\sigma$.

For the perovskite cuprates, the main peak of Cu 2$p$ usually broadens due to the presence of the mixing of $2p^53d^{10}L$ final states with $d^9L$ configurations on neighboring octahedra when the $d^9$ and $d^9L$ initial states are admixed.
On the other hand, for Ba$_3$CuSb$_2$O$_9$, the transition from $d^9L$ to $2p^53d^{10}L$ appears around 1-2 eV higher than the main peak with a reasonable parameter set as shown in Fig. S3(a), indicating that the average concentration of $\sim$ 1/6 holes per Cu fits the experimental spectra.
As for the O 1$s$ spectra [Fig. S3(b)], the prepeak or transition from $d^9L$ to $cd^9$ [so-called Zhang-Rice singlet (ZRS) feature] appears about 2-3 eV lower than the upper Hubbard band (UHB) peak (the transition to the O 2$p$ component mixed into the unoccupied Cu 3$d$ level), and is qualitatively consistent with experimental results.


\begin{figure}
	\centering
	\includegraphics[width=0.8\linewidth]{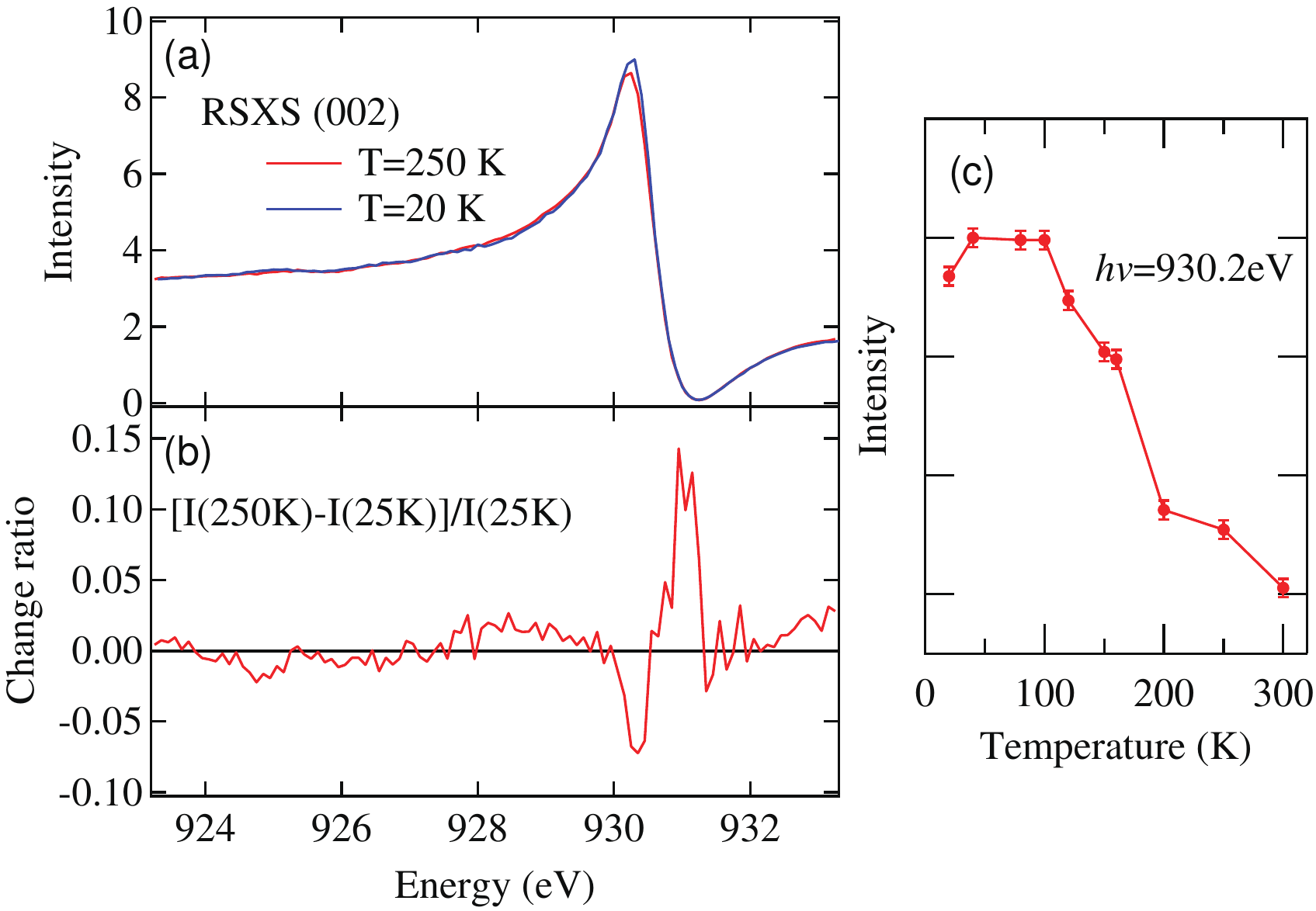}
	\caption{
		Temperature dependence of the static RSXS for hexagonal Ba$_3$CuSb$_2$O$_9$ at Q=(002) near the Cu $L_3$ edge.
		The RSXS intensity was normalized by the intensity of XAS at $hv$=923 eV in the total fluorescence yield mode which was simultaneously measured with RSXS. (a) RSXS spectra. (b) Change ratio of (a); $[I(250\mathrm K)-I(25\mathrm K)]/I(25\mathrm K)$. (c) Temperature dependence of RSXS at $hv$=930.2 eV.
	}
\end{figure}

\begin{figure}
	\centering
	\includegraphics[width=0.6\linewidth]{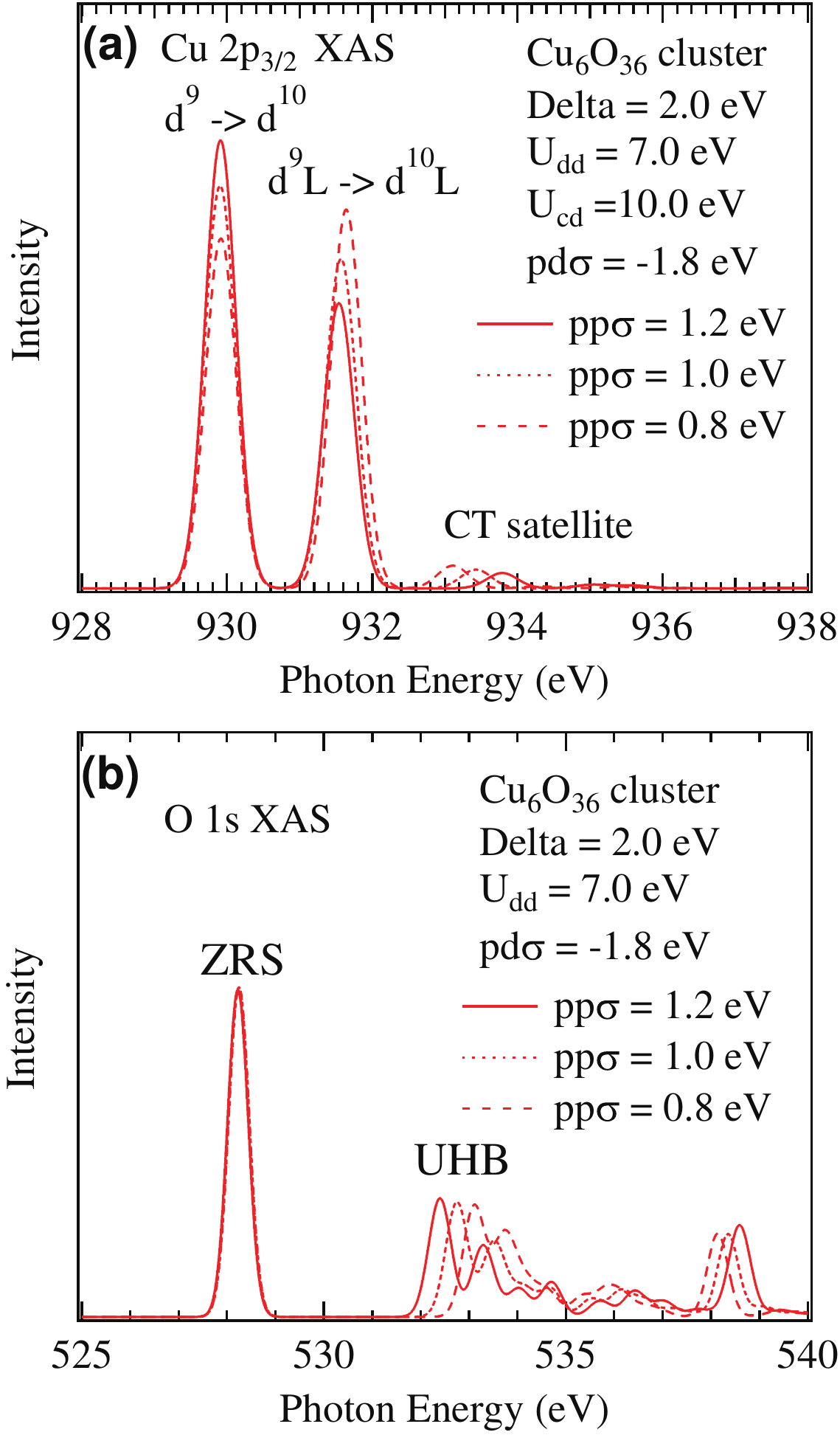}
	\caption{
		Configuration interaction calculation for Cu$_6$O$_{36}$ cluster with one oxygen hole of (a) Cu 2p$_{3/2}$ and (b) O 1$s$.
	}
\end{figure}


\clearpage

\begin{quote}

\end{quote}